\documentclass[sigplan, 10pt]{acmart}
\renewcommand\footnotetextcopyrightpermission[1]{} % removes footnote with conference information in first column
\setcopyright{none}
\usepackage{booktabs}
\usepackage{subcaption}

\newcommand{\code}[1]{\texttt{#1}}
\usepackage{balance}
\usepackage{acronym}

\usepackage{xspace}
\newcommand{\sys}{HyperProv\xspace}

\acrodef{uit}[UiT]{University of Tromsø}
\acrodef{api}[API]{Application Programming Interface} %\glsunset{api}
\acrodef{pow}[PoW]{Proof-of-Work}
\acrodef{pos}[PoS]{Proof-of-Stake}
\acrodef{grpc}[gRPC]{gRPC Remote Procedure Calls}
\acrodef{ca}[CA]{Certificate Authority}
\acrodef{dht}[DHT]{Distributed Hash Table}
\acrodef{p2p}[P2P]{Peer-to-peer}
\acrodef{opm}[OPM]{Open Provenance Model}
\acrodef{puf}[PUF]{Physical Unclonable Functions}
\acrodef{hlf}[HLF]{Hyperledger Fabric}
\acrodef{sdk}[SDK]{Software Development Kit}
\acrodef{npm}[NPM]{Node.js Package Manager}
\acrodef{gsod}[GSOD]{Global Surface Summary of the Day}
\acrodef{dag}[DAG]{directed acyclic graph}
\acrodef{gcp}[GCP]{Google Cloud Platform}
\acrodef{crdt}[CRDT]{Conflict-free Replicated Data Types}
\acrodef{tls}[TLS]{Transport Layer Security}
\acrodef{evm}[EVM]{Ethereum Virtual Machine}
\acrodef{rpi}[RPi]{Raspberry Pi}
\acrodef{iot}[IoT]{Internet of Things}

\begin{document}
%
%\bstctlcite{IEEEexample:BSTcontrol} % commands to IEEE BST file


%% Title information
\title[HyperProv]{HyperProv: Decentralized Resilient Data Provenance at the Edge with Blockchains}         %% [Short Title] is optional;
% \title[HyperProv]{Poster Abstract: HyperProv: Blockchain-based Data Provenance using Hyperledger Fabric}         %% [Short Title] is optional;
                                        %% when present, will be used in
                                        %% header instead of Full Title.
% \titlenote{with title note}             %% \titlenote is optional;
                                        %% can be repeated if necessary;
                                        %% contents suppressed with 'anonymous'
% \subtitle{Subtitle}                     %% \subtitle is optional
% \subtitlenote{with subtitle note}       %% \subtitlenote is optional;
                                        %% can be repeated if necessary;
                                        %% contents suppressed with 'anonymous'
%% Author information
%% Contents and number of authors suppressed with 'anonymous'.
%% Each author should be introduced by \author, followed by
%% \authornote (optional), \orcid (optional), \affiliation, and
%% \email.
%% An author may have multiple affiliations and/or emails; repeat the
%% appropriate command.
%% Many elements are not rendered, but should be provided for metadata
%% extraction tools.

\author{Petter Tunstad}
\affiliation{
  \department{Department of Computer Science}
  \institution{UiT The Arctic University of Norway}
  \city{Troms\o}
  \country{Norway}
}
% \email{ptu001@post.uit.no}
\email{pettertunstad@gmail.com}

\author{Amin M. Khan}
\affiliation{
  \department{Department of Computer Science}
  \institution{UiT The Arctic University of Norway}
  \city{Troms\o}
  \country{Norway}
}
\email{amin.khan@uit.no}

\author{Phuong Hoai Ha}
\affiliation{
  \department{Department of Computer Science}
  \institution{UiT The Arctic University of Norway}
  \city{Troms\o}
  \country{Norway}
}
\email{phuong.hoai.ha@uit.no}

% ABSTRACT
% !TeX encoding = UTF-8
% !TeX spellcheck = en_GB

%% Abstract
%% Note: \begin{abstract}...\end{abstract} environment must come
%% before \maketitle command

% \renewcommand{\sys}{HyperProv\xspace}

\begin{abstract}

Data provenance and lineage are critical for ensuring integrity and reproducibility of information in research and application.
This is particularly challenging for distributed scenarios, where data may be originating from decentralized sources without any central control by a single trusted entity.
We present \sys{}, a general framework for data provenance based on the permissioned blockchain Hyperledger Fabric (HLF),
and to the best of our knowledge, the first system that is ported to ARM based devices such as Raspberry Pi (RPi).
\sys{} tracks the metadata, operation history and data lineage 
through a set of built-in queries using smart contracts, enabling lightweight retrieval of provenance data.
\sys{} provides convenient integration through a NodeJS client library, and also includes off-chain storage through the SSH file system. 
We evaluate \sys{}'s performance, throughput, resource consumption, and energy efficiency on x86-64 machines, as well as on RPi devices for IoT use cases at the edge.

\end{abstract}

%% 2012 ACM Computing Classification System (CSS) concepts
%% Generate at 'http://dl.acm.org/ccs/ccs.cfm'.
\begin{CCSXML}
<ccs2012>
<concept>
<concept_id>10010520.10010521.10010537</concept_id>
<concept_desc>Computer systems organization~Distributed architectures</concept_desc>
<concept_significance>500</concept_significance>
</concept>
</ccs2012>
\end{CCSXML}

\ccsdesc[500]{Computer systems organization~Distributed architectures}
%% End of generated code

%% Keywords
%% comma separated list
\keywords{
Blockchain,
Data Provenance,
Edge Computing
}  %% \keywords are mandatory in final camera-ready submission

% HOW TO:
% 
% To copy text without line breaks, copy from README.md, or
% Comment the \begin{abstract} and \end{abstract}
% Run the following command:
% pandoc Sections/abstract.tex -s -o Sections/abstract.html
% Copy the text from page in the browser

\maketitle
\pagestyle{plain} % removes running headers
\thispagestyle{empty} % removes page number from first page

%% *** Introduction
% !TeX encoding = UTF-8
% !TeX spellcheck = en_US

% Section: INTRODUCTION
\section{Introduction}
\label{sec:introduction}

Over the last decades, the size of data utilized in research has increased significantly,
highlighting the importance of data provenance systems~\cite{surveryOnProvenance, provenanceInEscience,muniswamy2010provenance}
in order to ensure the quality and integrity of the information, and to counteract accidental or malicious data manipulation and corruption.
We present \sys{}, a permissioned blockchain based provenance system using \ac{hlf}~\cite{Androulaki2018HLF} to provide guarantees for provenance and lineage 
of data by storing the provenance metadata in a tamper-proof ledger. 
We consider the use case of \ac{iot} data at the edge to demonstrate utility and applicability of \sys{}.
We evaluate in detail~\cite{tunstad2019hyperprov} throughput, resource consumption and energy efficiency of \sys{} both on x86-64 commodity hardware and \ac{rpi} ARM64 devices,
in order to compare its performance with other recent blockchain-based systems~\cite{provchain, smartprovenance}.

To the best of our knowledge, we are the first to run a provenance system featuring \ac{hlf}'s first long term release,
and are also the first one to run provenance system based on \ac{hlf} on ARM devices.
Our contributions in porting HLF for ARM devices have already generated significant uptake and recognition in the community, 
with more than 500 downloads in the first 2-3 months%
\footnote{\url{https://github.com/Tunstad/Hyperprov}}$^{,}$%
\footnote{\url{https://hub.docker.com/u/ptunstad}}.
We believe \sys{} demonstrates the feasibility of using edge devices such as \ac{rpi} for data provenance through blockchains.
We hope that by building and releasing Docker images for ARM, we can pave way for other innovative solutions employing \ac{hlf} for edge computing. 
\sys{}'s NodeJS client library hides away the complexity of working with \ac{hlf}, 
allowing for easily plugging in \sys{} with other domain-specific data provenance systems.

% *** Section: Related Work
% !TeX encoding = UTF-8
% !TeX spellcheck = en_US

\section{Related Work}
\label{sec__related_work}

Many projects have tackled domain-specific as well as general purpose data provenance,
such as
	% PAAS,
	% Chimera,
	% MyGrid,
	% CMCS,
	% ESSW,
	% and Trio,
	PAAS~\cite{muniswamy2010provenance},
	Chimera~\cite{chimera},
	MyGrid~\cite{mygrid},
	CMCS~\cite{cmcs},
	ESSW~\cite{essw},
	and Trio~\cite{trio},
among others~\cite{surveryOnProvenance, provenanceInEscience}.
We limit our focus here on recent data provenance systems that employ blockchains~\cite{smartprovenance, moskowHLprov, provchain}, and
\sys{} is distinguished from them as it uses permissioned blockchains like \ac{hlf} that have much less resource requirements compared to public blockchains~\cite{thakkar2018performance}, 
limits recording only provenance metadata in the blockchain while moving actual data to off-chain storage, and demonstrates practical applications of using devices such as \ac{rpi} at the edge without relying on constant connectivity to the cloud.
With respect to \sys{}'s implementation, 
other recent works not focused on provenance but related to \ac{hlf} and \ac{rpi}
are also of interest to us.
Vegvisir~\cite{vegvisir} handles network partitions,
which is critical to successful deployment in \ac{iot} and edge scenarios,
FastFabric~\cite{Gorenflo2019FastFabric} improves on the throughput of \ac{hlf},
while~\cite{Selimi2018Blockchain} demonstrates feasibility and utility of \ac{hlf}
for edge scenarios like wireless mesh networks.
We leave the detailed comparison to the technical report~\cite{tunstad2019hyperprov}.

% \input{Sections/hlf}

%% *** Section: Design & Architecture
% !TeX encoding = UTF-8
% !TeX spellcheck = en_US

% Section: System
\section{\sys{} Data Provenance System}
\label{sec__system}

\sys{} follows the features from the Open Provenance Model~\cite{opm},
while still supporting extensions to support domain-specific provenance metadadata.
To this end, \sys{} stores the checksum, editors, operations, data ownership, and data pointers in the blockchain,
while delegating the storage of actual data to a separate \emph{off-chain} storage system.
\sys{} consists of 
\ac{hlf}-framework running in Docker containers
and the NodeJS based client library, 
and is supplemented by an off-chain storage system built using SSHFS.
The goal of the system is to enable seamless storage of provenance metadata in a tamper-proof blockchain framework when accessing and storing data in a \emph{pluggable} storage service.
Peer nodes are responsible for hosting the chaincode, which is the executable logic that can append or query data stored in the ledger. 
The chaincode consists of a few functions that are mirrored and available across all peer nodes. 
The core data currently stored in the blockchain is the checksum of every data item, the data location, a certificate pertaining to who stored the data, a list of other data items that were used to create an item, and a custom field for any additional metadata.
%
% \subsection{Exposed API}
The \sys{}'s client library enables the use of the \sys{} system for a wide range of functionality with only a few limited operators, such as \code{Init}, \code{Post},\code{Get}, \code{StoreData}, \code{GetData}, etc., see~\cite{tunstad2019hyperprov} for the details.

% \input{Sections/architecture}

%% *** Section: Implementation
% \input{Sections/implementation}

% *** Section: Evaluation
% !TeX encoding = UTF-8
% !TeX spellcheck = en_US

% Section: EVALUATION
\section{Evaluation}
\label{sec__evaluation}

\begin{figure}[tbp]
\centering
\includegraphics[width=0.95\columnwidth]{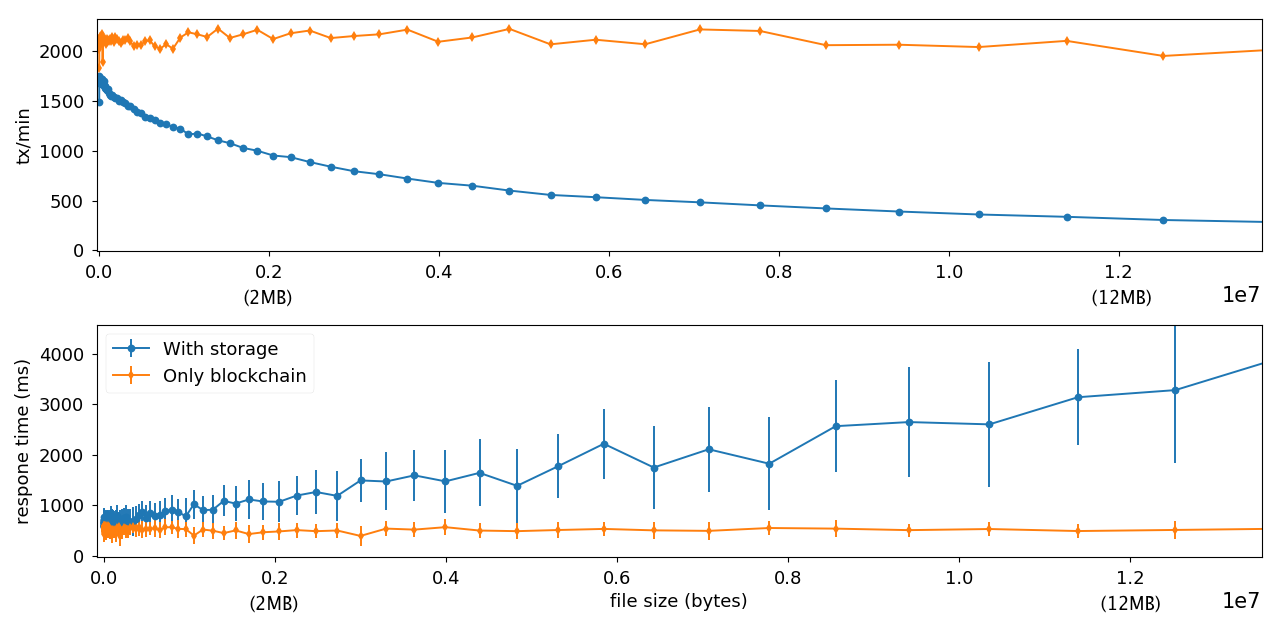}
\caption{Throughput and response times for desktop}
\label{fig:throughput-desktop}
\end{figure}

We perform extensive evaluation~\cite{tunstad2019hyperprov} using two different setups of the same network. 
The first consists of desktop nodes and the other consists of \ac{rpi} devices. 
The desktop setup has 4 machines:
    2 Intel Xeon E5-1603 CPU 2.80GHz, 
    1 Intel Core i7-4700MQ CPU 2.40GHz,
    and 1 Intel Corei3-2310M CPU 2.10GHz. 
All have Ubuntu 16.04 OS and are equipped with SSD storage. 
Each of the four nodes run peer docker containers, 
whereas one Xeon machine runs the orderer, and all use the official \ac{hlf} docker images.
The second setup consists of 4 ARM-based \ac{rpi} 3B+ 1.4GHz Cortex-A53 devices, interconnected on the same network switch. 
The \ac{rpi} runs the unofficial RaspberryPi 3 Debian Buster 64-bit OS, since the newer \ac{hlf} versions require 64-bit support. 
Due to the lack of other supported docker images, we have compiled our own images for the ARM64 architecture for \ac{rpi}.
Off-chain storage component based on SSH file system always runs on a separate node.
The measurements are performed using our custom benchmarking program. 
% for both the 2.8GHz Xeon E5 CPU for desktop, and the 1.4GHz (throttled at 600MHz) Cortex-A53 CPU of the \ac{rpi}. 
The energy consumption is measured using an ODROID V3 power meter placed between the device and the power source.

\begin{figure}[tbp]
	\centering
	\includegraphics[width=0.95\columnwidth]{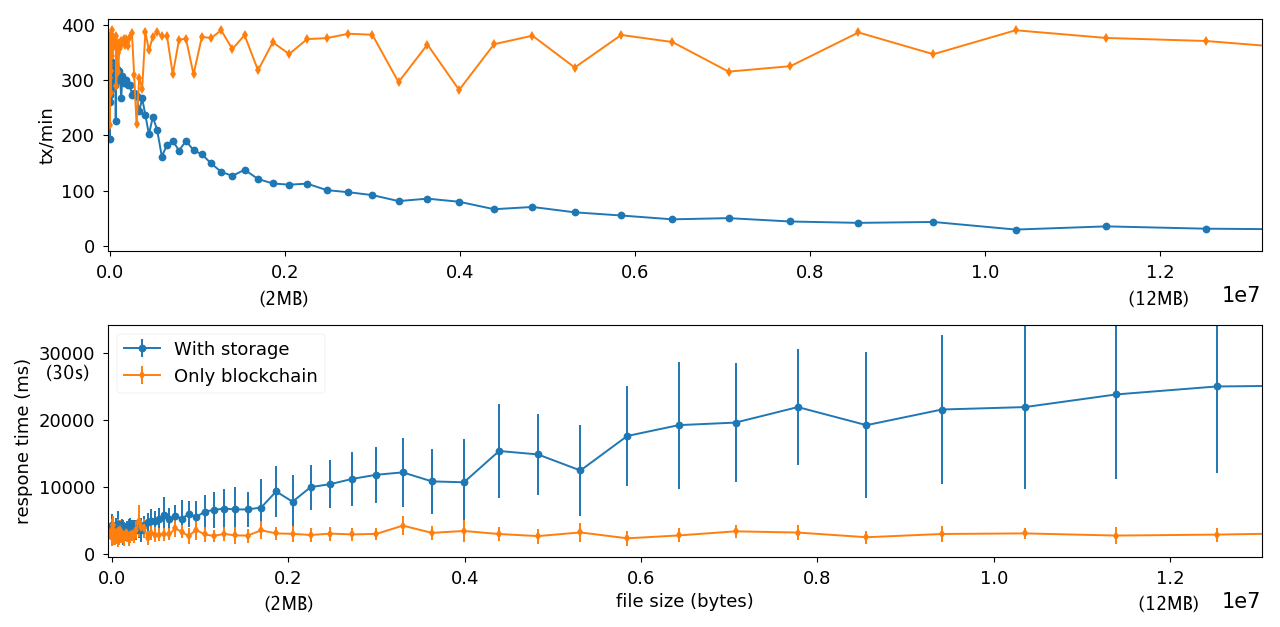}
	\caption{Throughput and response times for \ac{rpi}}
	\label{fig:throughput-rpi}
\end{figure}

\begin{figure}[tbp]
	\centering
	\includegraphics[width=0.85\columnwidth]{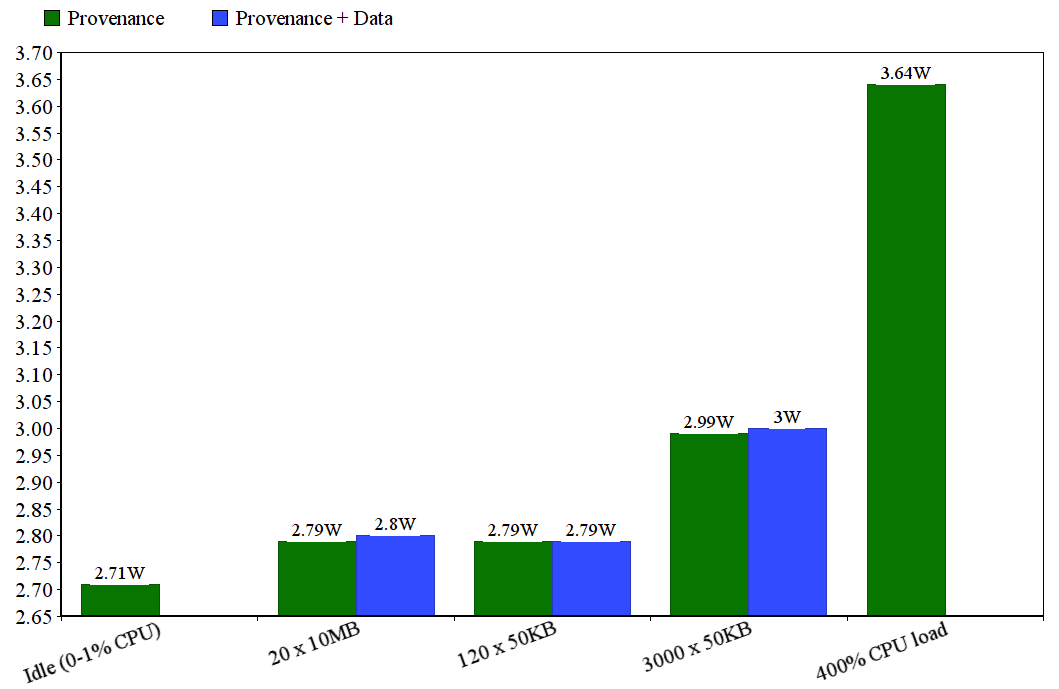}
	\caption{Energy consumption on \ac{rpi}, 10-minute intervals}
	\label{fig:energy-rpi}
\end{figure}

Fig.~\ref{fig:throughput-desktop} shows how increasing the size of data items impacts both throughput and response times, when off-chain storage is involved for desktop machines which incurs the overhead of data transfer and checksum calculation.
Fig.~\ref{fig:throughput-rpi} shows similar trend for throughput and response times for \ac{rpi} though greater variation, however absolute performance for \ac{rpi} is lower than desktop machines as expected owing to the limited hardware capacity.
Measurements of the energy consumption of \ac{rpi} devices running both peer and client processes for 10 minutes (as shown in Fig.~\ref{fig:energy-rpi}) 
highlight that running \sys{} without any active transactions barely consumes any power (2.71W) compared to an idle \ac{rpi} running without \ac{hlf},
while at the peak load level consumes only 10.7\% more as compared to idle, and maximum up to 3.64W.

% *** Section: Discussion
% \input{Sections/discussion} 
% \input{Sections/future_work} 
 
% *** Section: Conclusion
% \input{Sections/conclusion}

% *** Acknowledgement
% !TeX encoding = UTF-8
% !TeX spellcheck = en_GB

%% Acknowledgments
\begin{acks}                            %% acks environment is optional
                                        %% contents suppressed with 'anonymous'
  %% Commands \grantsponsor{<sponsorID>}{<name>}{<url>} and
  %% \grantnum[<url>]{<sponsorID>}{<number>} should be used to
  %% acknowledge financial support and will be used by metadata
  %% extraction tools.
%   This material is based upon work supported by the
%   \grantsponsor{GS100000001}{National Science
%     Foundation}{http://dx.doi.org/10.13039/100000001} under Grant
%   No.~\grantnum{GS100000001}{nnnnnnn} and Grant
%   No.~\grantnum{GS100000001}{mmmmmmm}.  Any opinions, findings, and
%   conclusions or recommendations expressed in this material are those
%   of the author and do not necessarily reflect the views of the
%   National Science Foundation.
  
This work is funded by the  \grantsponsor{}{Research Council of Norway}{https://www.forskningsradet.no/en/}
under PREAPP project (grant n$^{\circ}$~~\grantnum{}{231746}) 
and eX3 project (grant n$^{\circ}$~\grantnum{}{270053}).
\end{acks}

% \vspace{-10pt}
% \clearpage

% *** References section
% \bibliographystyle{\bibstylefile}
\balance
\bibliography{main}


\begin{thebibliography}{18}

%%% the Web address.

\ifx \showCODEN    \undefined \def \showCODEN     #1{\unskip}     \fi
\ifx \showDOI      \undefined \def \showDOI       #1{#1}\fi
\ifx \showISBNx    \undefined \def \showISBNx     #1{\unskip}     \fi
\ifx \showISBNxiii \undefined \def \showISBNxiii  #1{\unskip}     \fi
\ifx \showISSN     \undefined \def \showISSN      #1{\unskip}     \fi
\ifx \showLCCN     \undefined \def \showLCCN      #1{\unskip}     \fi
\ifx \shownote     \undefined \def \shownote      #1{#1}          \fi
\ifx \showarticletitle \undefined \def \showarticletitle #1{#1}   \fi
\ifx \showURL      \undefined \def \showURL       {\relax}        \fi
% invisible to TeX
\providecommand\bibfield[2]{#2}
\providecommand\bibinfo[2]{#2}
\providecommand\natexlab[1]{#1}
\providecommand\showeprint[2][]{arXiv:#2}

\bibitem[\protect\citeauthoryear{Androulaki, Manevich, Muralidharan, Murthy,
  Nguyen, Sethi, Singh, Smith, Sorniotti, Stathakopoulou, Vukoli{\'{c}},
  Barger, Cocco, Yellick, Bortnikov, Cachin, Christidis, {De Caro}, Enyeart,
  Ferris, and Laventman}{Androulaki et~al\mbox{.}}{2018}]%
        {Androulaki2018HLF}
\bibfield{author}{\bibinfo{person}{Elli Androulaki}, \bibinfo{person}{Yacov
  Manevich}, \bibinfo{person}{Srinivasan Muralidharan}, \bibinfo{person}{Chet
  Murthy}, \bibinfo{person}{Binh Nguyen}, \bibinfo{person}{Manish Sethi},
  \bibinfo{person}{Gari Singh}, \bibinfo{person}{Keith Smith},
  \bibinfo{person}{Alessandro Sorniotti}, \bibinfo{person}{Chrysoula
  Stathakopoulou}, \bibinfo{person}{Marko Vukoli{\'{c}}},
  \bibinfo{person}{Artem Barger}, \bibinfo{person}{Sharon~Weed Cocco},
  \bibinfo{person}{Jason Yellick}, \bibinfo{person}{Vita Bortnikov},
  \bibinfo{person}{Christian Cachin}, \bibinfo{person}{Konstantinos
  Christidis}, \bibinfo{person}{Angelo {De Caro}}, \bibinfo{person}{David
  Enyeart}, \bibinfo{person}{Christopher Ferris}, {and}
  \bibinfo{person}{Gennady Laventman}.} \bibinfo{year}{2018}\natexlab{}.
\newblock \showarticletitle{{Hyperledger Fabric: A Distributed Operating System
  for Permissioned Blockchains}}. In \bibinfo{booktitle}{\emph{EuroSys '18}}.
\newblock
\showISBNx{9781450355841}
\urldef\tempurl%
\url{https://doi.org/10.1145/3190508.3190538}
\showDOI{\tempurl}


\bibitem[\protect\citeauthoryear{{Cui} and {Widom}}{{Cui} and {Widom}}{2000}]%
        {trio}
\bibfield{author}{\bibinfo{person}{Y. {Cui}} {and} \bibinfo{person}{J.
  {Widom}}.} \bibinfo{year}{2000}\natexlab{}.
\newblock \showarticletitle{{Practical lineage tracing in data warehouses}}. In
  \bibinfo{booktitle}{\emph{{Proceedings of 16th International Conference on
  Data Engineering (Cat. No.00CB37073)}}}. \bibinfo{pages}{367--378}.
\newblock
\showISSN{1063-6382}
\urldef\tempurl%
\url{https://doi.org/10.1109/ICDE.2000.839437}
\showDOI{\tempurl}


\bibitem[\protect\citeauthoryear{Demichev, Kryukov, and Prikhodko}{Demichev
  et~al\mbox{.}}{2018}]%
        {moskowHLprov}
\bibfield{author}{\bibinfo{person}{Andrey Demichev}, \bibinfo{person}{Alexander
  Kryukov}, {and} \bibinfo{person}{Nikolai Prikhodko}.}
  \bibinfo{year}{2018}\natexlab{}.
\newblock \showarticletitle{{The Approach to Managing Provenance Metadata and
  Data Access Rights in Distributed Storage Using the Hyperledger Blockchain
  Platform}}.
\newblock \bibinfo{journal}{\emph{CoRR}}  \bibinfo{volume}{abs/1811.12706}
  (\bibinfo{year}{2018}).
\newblock


\bibitem[\protect\citeauthoryear{{Foster}, {Vockler}, {Wilde}, and {Yong
  Zhao}}{{Foster} et~al\mbox{.}}{2002}]%
        {chimera}
\bibfield{author}{\bibinfo{person}{I. {Foster}}, \bibinfo{person}{J.
  {Vockler}}, \bibinfo{person}{M. {Wilde}}, {and} \bibinfo{person}{{Yong
  Zhao}}.} \bibinfo{year}{2002}\natexlab{}.
\newblock \showarticletitle{{Chimera: a virtual data system for representing,
  querying, and automating data derivation}}. In
  \bibinfo{booktitle}{\emph{{Proceedings 14th International Conference on
  Scientific and Statistical Database Management}}}. \bibinfo{pages}{37--46}.
\newblock
\showISSN{1099-3371}
\urldef\tempurl%
\url{https://doi.org/10.1109/SSDM.2002.1029704}
\showDOI{\tempurl}


\bibitem[\protect\citeauthoryear{{Frew} and {Bose}}{{Frew} and {Bose}}{2001}]%
        {essw}
\bibfield{author}{\bibinfo{person}{J. {Frew}} {and} \bibinfo{person}{R.
  {Bose}}.} \bibinfo{year}{2001}\natexlab{}.
\newblock \showarticletitle{{Earth System Science Workbench: a data management
  infrastructure for earth science products}}. In
  \bibinfo{booktitle}{\emph{{Proceedings Thirteenth International Conference on
  Scientific and Statistical Database Management. SSDBM 2001}}}.
  \bibinfo{pages}{180--189}.
\newblock
\showISSN{1099-3371}
\urldef\tempurl%
\url{https://doi.org/10.1109/SSDM.2001.938550}
\showDOI{\tempurl}


\bibitem[\protect\citeauthoryear{Gorenflo, Lee, Golab, and Keshav}{Gorenflo
  et~al\mbox{.}}{2019}]%
        {Gorenflo2019FastFabric}
\bibfield{author}{\bibinfo{person}{Christian Gorenflo},
  \bibinfo{person}{Stephen Lee}, \bibinfo{person}{Lukasz Golab}, {and}
  \bibinfo{person}{S. Keshav}.} \bibinfo{year}{2019}\natexlab{}.
\newblock \showarticletitle{{FastFabric: Scaling Hyperledger Fabric to 20,000
  Transactions per Second}}. In \bibinfo{booktitle}{\emph{IEEE International
  Conference on Blockchain and Cryptocurrency (ICBC '19)}}.
  \bibinfo{publisher}{IEEE}.
\newblock


\bibitem[\protect\citeauthoryear{Herschel, Diestelk\"{a}mper, and
  Ben~Lahmar}{Herschel et~al\mbox{.}}{2017}]%
        {surveryOnProvenance}
\bibfield{author}{\bibinfo{person}{Melanie Herschel}, \bibinfo{person}{Ralf
  Diestelk\"{a}mper}, {and} \bibinfo{person}{Houssem Ben~Lahmar}.}
  \bibinfo{year}{2017}\natexlab{}.
\newblock \showarticletitle{{A Survey on Provenance: What for? What Form? What
  from?}}
\newblock \bibinfo{journal}{\emph{The VLDB Journal}} \bibinfo{volume}{26},
  \bibinfo{number}{6} (\bibinfo{date}{Dec.} \bibinfo{year}{2017}),
  \bibinfo{pages}{881--906}.
\newblock
\showISSN{1066-8888}
\urldef\tempurl%
\url{https://doi.org/10.1007/s00778-017-0486-1}
\showDOI{\tempurl}


\bibitem[\protect\citeauthoryear{{Karlsson}, {Jiang}, {Wicker}, {Adams}, {Ma},
  {van Renesse}, and {Weatherspoon}}{{Karlsson} et~al\mbox{.}}{2018}]%
        {vegvisir}
\bibfield{author}{\bibinfo{person}{K. {Karlsson}}, \bibinfo{person}{W.
  {Jiang}}, \bibinfo{person}{S. {Wicker}}, \bibinfo{person}{D. {Adams}},
  \bibinfo{person}{E. {Ma}}, \bibinfo{person}{R. {van Renesse}}, {and}
  \bibinfo{person}{H. {Weatherspoon}}.} \bibinfo{year}{2018}\natexlab{}.
\newblock \showarticletitle{{Vegvisir: A Partition-Tolerant Blockchain for the
  Internet-of-Things}}. In \bibinfo{booktitle}{\emph{{2018 IEEE 38th
  International Conference on Distributed Computing Systems (ICDCS)}}}.
  \bibinfo{pages}{1150--1158}.
\newblock
\showISSN{2575-8411}
\urldef\tempurl%
\url{https://doi.org/10.1109/ICDCS.2018.00114}
\showDOI{\tempurl}


\bibitem[\protect\citeauthoryear{{Liang}, {Shetty}, {Tosh}, {Kamhoua}, {Kwiat},
  and {Njilla}}{{Liang} et~al\mbox{.}}{2017}]%
        {provchain}
\bibfield{author}{\bibinfo{person}{X. {Liang}}, \bibinfo{person}{S. {Shetty}},
  \bibinfo{person}{D. {Tosh}}, \bibinfo{person}{C. {Kamhoua}},
  \bibinfo{person}{K. {Kwiat}}, {and} \bibinfo{person}{L. {Njilla}}.}
  \bibinfo{year}{2017}\natexlab{}.
\newblock \showarticletitle{{ProvChain: A Blockchain-Based Data Provenance
  Architecture in Cloud Environment with Enhanced Privacy and Availability}}.
  In \bibinfo{booktitle}{\emph{{2017 17th IEEE/ACM International Symposium on
  Cluster, Cloud and Grid Computing (CCGRID)}}}. \bibinfo{pages}{468--477}.
\newblock
\urldef\tempurl%
\url{https://doi.org/10.1109/CCGRID.2017.8}
\showDOI{\tempurl}


\bibitem[\protect\citeauthoryear{Moreau, Clifford, Freire, Futrelle, Gil,
  Groth, Kwasnikowska, Miles, Missier, Myers, Plale, Simmhan, Stephan, and den
  Bussche}{Moreau et~al\mbox{.}}{2011}]%
        {opm}
\bibfield{author}{\bibinfo{person}{Luc Moreau}, \bibinfo{person}{Ben Clifford},
  \bibinfo{person}{Juliana Freire}, \bibinfo{person}{Joe Futrelle},
  \bibinfo{person}{Yolanda Gil}, \bibinfo{person}{Paul Groth},
  \bibinfo{person}{Natalia Kwasnikowska}, \bibinfo{person}{Simon Miles},
  \bibinfo{person}{Paolo Missier}, \bibinfo{person}{Jim Myers},
  \bibinfo{person}{Beth Plale}, \bibinfo{person}{Yogesh Simmhan},
  \bibinfo{person}{Eric Stephan}, {and} \bibinfo{person}{Jan~Van den Bussche}.}
  \bibinfo{year}{2011}\natexlab{}.
\newblock \showarticletitle{{The Open Provenance Model core specification
  (v1.1)}}.
\newblock \bibinfo{journal}{\emph{Future Generation Computer Systems}}
  \bibinfo{volume}{27}, \bibinfo{number}{6} (\bibinfo{year}{2011}),
  \bibinfo{pages}{743 -- 756}.
\newblock
\showISSN{0167-739X}
\urldef\tempurl%
\url{https://doi.org/10.1016/j.future.2010.07.005}
\showDOI{\tempurl}


\bibitem[\protect\citeauthoryear{Muniswamy-Reddy, Macko, and
  Seltzer}{Muniswamy-Reddy et~al\mbox{.}}{2010}]%
        {muniswamy2010provenance}
\bibfield{author}{\bibinfo{person}{Kiran-Kumar Muniswamy-Reddy},
  \bibinfo{person}{Peter Macko}, {and} \bibinfo{person}{Margo~I Seltzer}.}
  \bibinfo{year}{2010}\natexlab{}.
\newblock \showarticletitle{{Provenance for the Cloud}}. In
  \bibinfo{booktitle}{\emph{USENIX FAST'10}}.
\newblock


\bibitem[\protect\citeauthoryear{Pancerella, Hewson, Koegler, Leahy, Lee, Rahn,
  Yang, Myers, Didier, McCoy, Schuchardt, Stephan, Windus, Amin, Bittner,
  Lansing, Minkoff, Nijsure, von Laszewski, Pinzon, Ruscic, Wagner, Wang, Pitz,
  Ho, Montoya, Xu, Allison, Green, and Frenklach}{Pancerella
  et~al\mbox{.}}{2003}]%
        {cmcs}
\bibfield{author}{\bibinfo{person}{Carmen Pancerella}, \bibinfo{person}{John
  Hewson}, \bibinfo{person}{Wendy Koegler}, \bibinfo{person}{David Leahy},
  \bibinfo{person}{Michael Lee}, \bibinfo{person}{Larry Rahn},
  \bibinfo{person}{Christine Yang}, \bibinfo{person}{James~D. Myers},
  \bibinfo{person}{Brett Didier}, \bibinfo{person}{Renata McCoy},
  \bibinfo{person}{Karen Schuchardt}, \bibinfo{person}{Eric Stephan},
  \bibinfo{person}{Theresa Windus}, \bibinfo{person}{Kaizar Amin},
  \bibinfo{person}{Sandra Bittner}, \bibinfo{person}{Carina Lansing},
  \bibinfo{person}{Michael Minkoff}, \bibinfo{person}{Sandeep Nijsure},
  \bibinfo{person}{Gregor von Laszewski}, \bibinfo{person}{Reinhardt Pinzon},
  \bibinfo{person}{Branko Ruscic}, \bibinfo{person}{Al Wagner},
  \bibinfo{person}{Baoshan Wang}, \bibinfo{person}{William Pitz},
  \bibinfo{person}{Yen-Ling Ho}, \bibinfo{person}{David Montoya},
  \bibinfo{person}{Lili Xu}, \bibinfo{person}{Thomas~C. Allison},
  \bibinfo{person}{William~H. Green, Jr.}, {and} \bibinfo{person}{Michael
  Frenklach}.} \bibinfo{year}{2003}\natexlab{}.
\newblock \showarticletitle{{Metadata in the Collaboratory for Multi-scale
  Chemical Science}}. In \bibinfo{booktitle}{\emph{{Proceedings of the 2003
  International Conference on Dublin Core and Metadata Applications: Supporting
  Communities of Discourse and Practice---Metadata Research \& Applications}}}
  \emph{(\bibinfo{series}{DCMI '03})}. \bibinfo{publisher}{Dublin Core Metadata
  Initiative}, Article \bibinfo{articleno}{13}, \bibinfo{numpages}{9}~pages.
\newblock
\showISBNx{0974530301}


\bibitem[\protect\citeauthoryear{Ramachandran and Kantarcioglu}{Ramachandran
  and Kantarcioglu}{2018}]%
        {smartprovenance}
\bibfield{author}{\bibinfo{person}{Aravind Ramachandran} {and}
  \bibinfo{person}{Murat Kantarcioglu}.} \bibinfo{year}{2018}\natexlab{}.
\newblock \showarticletitle{{SmartProvenance: A Distributed, Blockchain Based
  DataProvenance System}}. In \bibinfo{booktitle}{\emph{CODASPY '18}}.
  \bibinfo{publisher}{ACM}, \bibinfo{pages}{35--42}.
\newblock
\showISBNx{978-1-4503-5632-9}
\urldef\tempurl%
\url{https://doi.org/10.1145/3176258.3176333}
\showDOI{\tempurl}


\bibitem[\protect\citeauthoryear{Selimi, Kabbinale, Ali, Navarro, and
  Sathiaseelan}{Selimi et~al\mbox{.}}{2018}]%
        {Selimi2018Blockchain}
\bibfield{author}{\bibinfo{person}{Mennan Selimi},
  \bibinfo{person}{Aniruddh~Rao Kabbinale}, \bibinfo{person}{Anwaar Ali},
  \bibinfo{person}{Leandro Navarro}, {and} \bibinfo{person}{Arjuna
  Sathiaseelan}.} \bibinfo{year}{2018}\natexlab{}.
\newblock \showarticletitle{{Towards Blockchain-enabled Wireless Mesh
  Networks}}. In \bibinfo{booktitle}{\emph{CryBlock '18}}.
  \bibinfo{publisher}{ACM Press}, \bibinfo{address}{Munich, Germany},
  \bibinfo{pages}{13--18}.
\newblock
\urldef\tempurl%
\url{https://doi.org/10.1145/3211933.3211936}
\showDOI{\tempurl}


\bibitem[\protect\citeauthoryear{Simmhan, Plale, and Gannon}{Simmhan
  et~al\mbox{.}}{2005}]%
        {provenanceInEscience}
\bibfield{author}{\bibinfo{person}{Yogesh~L. Simmhan}, \bibinfo{person}{Beth
  Plale}, {and} \bibinfo{person}{Dennis Gannon}.}
  \bibinfo{year}{2005}\natexlab{}.
\newblock \showarticletitle{{A Survey of Data Provenance in e-Science}}.
\newblock \bibinfo{journal}{\emph{SIGMOD Rec.}} \bibinfo{volume}{34},
  \bibinfo{number}{3} (\bibinfo{date}{Sept.} \bibinfo{year}{2005}),
  \bibinfo{pages}{31--36}.
\newblock
\showISSN{0163-5808}
\urldef\tempurl%
\url{https://doi.org/10.1145/1084805.1084812}
\showDOI{\tempurl}


\bibitem[\protect\citeauthoryear{Thakkar, Nathan, and Viswanathan}{Thakkar
  et~al\mbox{.}}{2018}]%
        {thakkar2018performance}
\bibfield{author}{\bibinfo{person}{Parth Thakkar}, \bibinfo{person}{Senthil
  Nathan}, {and} \bibinfo{person}{Balaji Viswanathan}.}
  \bibinfo{year}{2018}\natexlab{}.
\newblock \showarticletitle{{Performance Benchmarking and Optimizing
  Hyperledger Fabric Blockchain Platform}}. In
  \bibinfo{booktitle}{\emph{MASCOTS '18}}. \bibinfo{pages}{264--276}.
\newblock
\showISSN{2375-0227}
\urldef\tempurl%
\url{https://doi.org/10.1109/MASCOTS.2018.00034}
\showDOI{\tempurl}


\bibitem[\protect\citeauthoryear{Tunstad}{Tunstad}{2019}]%
        {tunstad2019hyperprov}
\bibfield{author}{\bibinfo{person}{Petter Tunstad}.}
  \bibinfo{year}{2019}\natexlab{}.
\newblock \emph{\bibinfo{title}{{Hyperprov: Blockchain-based Data Provenance
  using Hyperledger Fabric}}}.
\newblock \bibinfo{thesistype}{Master's\ thesis}. \bibinfo{school}{UiT The
  Arctic University of Norway}.
\newblock
\urldef\tempurl%
\url{https://hdl.handle.net/10037/15780}
\showURL{%
\tempurl}


\bibitem[\protect\citeauthoryear{Zhao, Goble, Stevens, and Bechhofer}{Zhao
  et~al\mbox{.}}{2004}]%
        {mygrid}
\bibfield{author}{\bibinfo{person}{Jun Zhao}, \bibinfo{person}{Carole Goble},
  \bibinfo{person}{Robert Stevens}, {and} \bibinfo{person}{Sean Bechhofer}.}
  \bibinfo{year}{2004}\natexlab{}.
\newblock \showarticletitle{{Semantically Linking and Browsing Provenance Logs
  for E-science}}. In \bibinfo{booktitle}{\emph{{Semantics of a Networked
  World. Semantics for Grid Databases}}},
  \bibfield{editor}{\bibinfo{person}{Mokrane Bouzeghoub},
  \bibinfo{person}{Carole Goble}, \bibinfo{person}{Vipul Kashyap}, {and}
  \bibinfo{person}{Stefano Spaccapietra}} (Eds.). \bibinfo{publisher}{Springer
  Berlin Heidelberg}, \bibinfo{address}{Berlin, Heidelberg},
  \bibinfo{pages}{158--176}.
\newblock
\showISBNx{978-3-540-30145-5}


\end{thebibliography}

%% Appendix
% \appendix
% \section{Appendix}

\end{document}